\documentclass[aps,twocolumn]{revtex4}
\usepackage{amsmath,amssymb}

\usepackage{epsfig}
\usepackage{graphicx}

\newcommand{\be}{\begin{equation}}
\newcommand{\ee}{\end{equation}}
\newcommand{\bea}{\begin{eqnarray}}
\newcommand{\eea}{\end{eqnarray}}
\newcommand{\nn}{\nonumber}

\begin{document}

\title{Conservation equation on braneworlds in six dimensions}

\author{Georgios Kofinas\footnote{gkofin@phys.uoa.gr; kofinas@ffn.ub.es}}

\date{\today}

%\address{~}

\address{Departament de F{\'\i}sica Fonamental,
Universitat de Barcelona, Diagonal 647, 08028 Barcelona, Spain}

\begin{abstract}

We study braneworlds in six-dimensional Einstein-Gauss-Bonnet
gravity. The Gauss-Bonnet term is crucial for the equations to be
well-posed in six dimensions when non-trivial matter on the brane is
included (the also involved induced gravity term is not significant
for their structure), and the matching conditions of the braneworld
are known. We show that the energy-momentum of the brane is always
conserved, independently of any regular bulk energy-momentum tensor,
contrary to the situation of the five-dimensional case.
\end{abstract}

\maketitle

Much work on braneworlds in six-dimensional spacetimes has been
done, especially during the last two years \cite{six}. It is known
that six-dimensional Einstein gravity cannot support a (thin)
braneworld with a non-trivial matter content different than a brane
tension \cite{cline}. The situation can be improved if a
Gauss-Bonnet term is added in the bulk action, in which case the
generic matching conditions of a 3-brane were derived in
\cite{ruth}. In the present note, we show that in such bulks the
energy momentum tensor of the brane is always conserved,
independently of the bulk energy-momentum tensor. On the contrary,
in five-dimensional bulks, as known, the conservation of
energy-momentum on the brane is violated when there is energy
transfer between brane and bulk, due to non-vanishing
transverse-parallel components of the bulk energy-momentum tensor.
\par
We consider the total gravitational brane-bulk action
 \bea &&
\!\!\!\!\!\!\!S_{gr}=\frac{1}{2\kappa_{6}^{2}}\!\int
\!d^6x\sqrt{\!-|\verb"g"|}
\left\{\mathcal{R}-2\Lambda_{6}+\alpha\,\Big(\mathcal{R}^{2}\!-4\mathcal{R}_{AB}\,
\mathcal{R}^{AB} \right.\nn
\\
&&\!\!\!\!\!\!\left.~{}+\mathcal{R}_{ABCD}\,\mathcal{R}^{ABCD}\Big)\!\right\}+
\frac{r_{c}^{2}}{2\kappa_{6}^{2}}\!\int \!d^4x\sqrt{\!-|g|}
\left(R-2\Lambda_{4}\right)\!, \label{6action}\eea where
calligraphic quantities refer to the bulk metric tensor $\verb"g"$,
while the regular ones to the brane metric tensor $g$. The
Gauss-Bonnet coupling $\alpha$ has dimensions $(length)^{2}$ and is
defined as
 \bea
\alpha=\frac{1}{ 8g_{s}^{2}}\,,
 \eea
with $g_{s}$ the string energy scale, while from the induced-gravity
crossover lenght scale $r_{c}$ we can define
 \bea
r_{c}=\frac{\kappa_6}{\kappa_4}=\frac{M_4}{M_6^2}\,.
 \label{distancescale}
 \eea
Here, $M_6$ is the fundamental six-dimensional Planck mass
$M_{6}^{-4}\!=\!\kappa_{6}^{2}\!=\!8\pi G_{6}$, while $M_{4}$ is
given by $M_{4}^{-2}\!=\!\kappa_{4}^{2}\!=\!8\pi G_{4}$. The brane
tension is
 \bea
\lambda={\Lambda_4 \over \kappa_4^2}\,.
 \eea
The field equations arising from the action (\ref{6action}) are \bea
&&\!\!\!\!\!\!\!\mathcal{G}_{AB}-\frac{\alpha}{2}(\mathcal{R}^2-4\mathcal{R}_{CD}\mathcal{R}^{CD}+\mathcal{R}_{CDEF}\mathcal{R}
^{CDEF})\verb"g"_{AB}+2\alpha\nn\\
&&\!\!\!\!\!\!\!\times(\mathcal{R}\mathcal{R}_{AB}\!-\!2\mathcal{R}_{AC}\mathcal{R}_{B}^{\,\,\,\,C}\!\!-\!
2\mathcal{R}_{ACBD}\mathcal{R}^{CD}\!\!+\!\mathcal{R}_{ACDE}\mathcal{R}_{B}^{\,\,\,\,CDE})\nn\\
&&\,\,\,\,\,\,\,\,\,\,\,\,\,\,\,\,\,\,\,
\,\,\,\,\,\,\,\,\,\,\,\,\,\,\,=\kappa_{6}^2
\mathcal{T}_{AB}-\Lambda_{6}
\verb"g"_{AB}+\kappa_{6}^2\,^{(loc)}\!T_{AB}\,\delta^{(2)},\label{6eqs}
\eea where $\mathcal{T}_{AB}$ is a regular bulk energy-momentum
tensor, $T_{AB}$ is the brane energy-momentum tensor,
$^{(loc)}\!T_{AB}=T_{AB}-\lambda
g_{AB}-(r_{c}^{2}/\kappa_{6}^2)G_{AB}$, and $\delta^{(2)}$ is the
two-dimensional delta function. Capital indices $A,B,...$ are
six-dimensional. Assuming that the bulk metric in the brane-adapted
coordinate system takes the axially symmetric form
 \bea
ds_{6}^2=dr^2+L^2(x,r)d\varphi^2+g_{\mu\nu}(x,r)dx^{\mu}dx^{\nu},\label{6metric}
\eea with $g_{\mu\nu}(x,0)$ being the braneworld metric and
$\varphi$ having the standard periodicity $2\pi$, under the usual
assumptions for conical singularities
$L(x,r)=\beta(x)r+\mathcal{O}(r^2)$ for $r\approx 0$,
$\partial_{r}L(x,0)=1$, $\partial_{r}g_{\mu\nu}(x,0)=0$, the general
matching conditions for imbedding the 3-brane in the six-dimensional
theory (\ref{6action}) were found in \cite{ruth} as follows \bea
&&\!\!\!\!\!\! K^{\alpha\lambda}_{\,\,\,\,\,\,\,\lambda}
K_{\alpha\mu\nu}\!-\!
K^{\alpha\lambda}_{\,\,\,\,\,\,\,\mu}K_{\alpha\nu\lambda}
\!+\!\frac{1}{2}(
K^{\alpha\lambda\sigma}K_{\alpha\lambda\sigma}\!-\!
K^{\alpha\lambda}_{\,\,\,\,\,\,\,\lambda}K_{\alpha\,\,\,\sigma}^{\,\,\,\sigma})g_{\mu\nu}\nn\\
&&\!\!\!\!\!\!\!+\Big(\!\beta^{-1}\!\!-\!1\!+\!\frac{r_{c}^2}{8\pi\alpha\beta}\!\Big)G_{\mu\nu}\!+\!
\frac{\kappa_{6}^{2}\lambda\!-\!2\pi(1\!-\!\beta)}{8\pi\alpha\beta}g_{\mu\nu}\!=\!
\frac{\kappa_{6}^{2}}{8\pi\alpha\beta}T_{\mu\nu}.
\nn\\
\label{matching} \eea  Here,
$K_{\alpha\mu\nu}=\verb"g"(\nabla_{\mu}n_{\alpha},\partial_{\nu})=n_{\alpha\mu;\nu}$
denote the extrinsic curvatures of the brane (symmetric in
$\mu,\nu$), where $n_{\alpha}$ ($\alpha=1,2$) are arbitrary unit
normals to the brane (indices $\alpha, \beta,...$ are lowered/raised
with the matrix
$\verb"g"_{\alpha\beta}=\verb"g"(n_{\alpha},n_{\beta})$ and its
inverse $\verb"g"^{\alpha\beta}$), while $\nabla$ (also denoted by
$;$) refers to the Christoffel connection of $\verb"g"$. For
extracting this singular part of equations (\ref{6eqs}), one has to
focus on the worst behaving pieces with the structure
$\delta(r)/L\sim \delta(r)/r$. Note that with respect to local
rotations $n_{\alpha}\rightarrow
O^{\,\,\,\beta}_{\alpha}(x^{A})\,n_{\beta}$,
$K_{\alpha\mu\nu}\rightarrow
O^{\,\,\,\beta}_{\alpha}K_{\beta\mu\nu}$ transforming as a vector,
thus Eq.(\ref{matching}) is invariant under changes of the normal
frame. It is also noticeable that the matching conditions
(\ref{matching}) are quadratic in the extrinsic curvature, while the
corresponding matching conditions of 5-dimensional Gauss-Bonnet
theory are cubic \cite{davis}. Focusing on the $\mathcal{O}(1/r)$
terms in the $r\mu$ components of equations (\ref{6eqs}) we obtain
the equation (correcting equation (16) of \cite{ruth}) \bea
&&\!\!\!\!\!\!\!
\mathcal{R}^{\!\alpha\sigma}_{\,\,\,\,\,\,\nu\sigma}K_{\alpha\,\,\,\lambda}^{\,\,\,\lambda}\!-\!
\mathcal{R}^{\!\alpha\sigma}_{\,\,\,\,\,\lambda\sigma}K_{\!\alpha\,\,\,\nu}^{\,\,\lambda}\!-\!
\mathcal{R}^{\!\alpha\lambda}_{\,\,\,\,\,\,\nu\sigma}K_{\!\alpha\,\,\,\lambda}^{\,\,\sigma}\!=
\frac{\beta_{,\mu}}{\beta}\!\Big[G_{\nu}^{\mu}\!-\!\frac{1}{4\alpha}\delta_{\nu}^{\mu}\nn\\
&&\!\!\!\!\!\!\!+K^{\alpha\sigma}_{\,\,\,\,\,\,\,\nu}
K_{\!\alpha\,\,\,\sigma}^{\,\,\mu}\!-\!
K^{\alpha\sigma}_{\,\,\,\,\,\,\,\sigma}K_{\alpha\,\,\,\nu}^{\,\,\,\mu}
\!+\!\frac{1}{2}\!(
K^{\alpha\sigma}_{\,\,\,\,\,\,\,\sigma}K_{\!\alpha\,\,\lambda}^{\,\,\lambda}
\!-\!K^{\alpha\sigma\lambda}\!K_{\alpha\sigma\lambda}\!)\delta_{\nu}^{\mu}\!\Big]\!.\nn\\
\label{christmas} \eea
\par
In higher codimensions, one has geometrical equations analogous to
those holding for hypersurfaces \cite{cap, carter}, namely \bea
\mathcal{R}_{\mu\nu\kappa\lambda}=R_{\mu\nu\kappa\lambda}+K^{\alpha}_{\,\,\,\mu\lambda}
K_{\alpha\nu\kappa}-K^{\alpha}_{\,\,\,\mu\kappa}K_{\alpha\nu\lambda}\,,
\label{GC} \eea \bea
\mathcal{R}^{\alpha}_{\,\,\,\mu\nu\lambda}=K^{\alpha}_{\,\,\,\mu\nu!\lambda}-K^{\alpha}_{\,\,\,\mu\lambda!\nu}\,,
\label{CM} \eea \bea \mathcal{R}^{\beta}_{\,\,\,\alpha
\mu\nu}=\Omega^{\beta}_{\,\,\,\alpha \mu\nu}+K_{\alpha
\mu}^{\,\,\,\,\,\,\lambda}K^{\beta}_{\,\,\,\nu\lambda}-K_{\alpha
\nu}^{\,\,\,\,\,\,\lambda}K^{\beta}_{\,\,\,\mu\lambda}\,.
\label{ricci} \eea The covariant derivative $!$ is defined with
respect to the connection $\varpi_{\beta\alpha
\mu}=\verb"g"(\nabla_{\mu}n_{\alpha},n_{\beta})$ as \bea
\Phi^{\alpha}_{\beta!\mu}=\Phi^{\alpha}_{\beta|\mu}+\varpi^{\alpha}_{\,\,\,\gamma\mu}\Phi^{\gamma}_{\beta}-
\varpi^{\gamma}_{\,\,\,\beta\mu}\Phi^{\alpha}_{\gamma},\label{new}
\eea for fields $\Phi^{\alpha}_{\beta}$ transforming as tensors
under normal frame rotations, $\Phi^{\alpha}_{\beta}\!\rightarrow\!
O_{\beta}^{\,\,\,\delta}(O^{-1})_{\gamma}^{\,\,\,\alpha}\Phi_{\delta}^{\gamma}$,
and $|$ refers to the Christoffel connection
$\gamma_{\mu\nu\lambda}\!=\!\verb"g"(\nabla_{\lambda}\partial_{\nu},\partial_{\mu})$
of the induced brane metric $g_{\mu\nu}$ [the derivative $|$ in
(\ref{new}) is meant on tangential indices $\mu,\nu,...$ that
$\Phi^{\alpha}_{\beta}$ may possess]. $\Omega^{\beta}_{\,\,\,\alpha
\mu\nu}$ is the curvature of the connection
$\varpi^{\beta}_{\,\,\,\alpha \mu}$, $\Omega^{\beta}_{\,\,\,\alpha
\mu\nu}=2\varpi^{\beta}_{\,\,\,\alpha[\nu,\mu]}+2\varpi^{\gamma}_{\,\,\,\alpha[\nu}\varpi^{\beta}
_{\,\,\,|\gamma|\mu]}$ \footnote{Analogous is our convention for the
ordering of indices in the definition of the other Riemann
tensors.}.
\par
From equations (\ref{matching}), differentiating with respect to $!$
and making use of the identity (\ref{CM}), we obtain
 \bea
&&\!\!\!\!\!\!\!\!\!\!\!\!\frac{\kappa_{6}^{2}}{8\pi\alpha\beta}T^{\mu}_{\nu|\mu}
=\mathcal{R}^{\alpha\mu}_{\,\,\,\,\,\,\,\nu\mu}K_{\alpha\,\,\,\lambda}^{\,\,\,\lambda}\!-\!
\mathcal{R}^{\alpha\mu}_{\,\,\,\,\,\,\lambda\mu}K_{\alpha\,\,\,\nu}^{\,\,\,\lambda}\!-\!
\mathcal{R}^{\alpha\lambda}_{\,\,\,\,\,\,\,\nu\mu}K_{\alpha\,\,\,\lambda}^{\,\,\,\mu}\nn\\
&&\,\,\,\,\,+\frac{\beta_{,\mu}}{\beta^2}\Big[\frac{\kappa_{6}^{2}}{8\pi\alpha}T^{\mu}_{\nu}
\!+\!\frac{2\pi\!-\!\kappa_{6}^{2}\lambda}{8\pi\alpha}\delta^{\mu}_{\nu}\!-\!\Big
(1\!+\!\frac{r_{c}^{2}}{8\pi\alpha}\Big) G^{\mu}_{\nu}\Big].
\label{con} \eea  It is noticeable that equation (\ref{con}) can
also arise from equations (\ref{matching}) and (\ref{GC}) without
using (\ref{CM}). Indeed, if
$S_{\mu\nu}=K^{\alpha\lambda}_{\,\,\,\,\,\,\,\lambda}
K_{\alpha\mu\nu}\!-\!
K^{\alpha\lambda}_{\,\,\,\,\,\,\,\mu}K_{\alpha\nu\lambda}$, the
matching conditions (\ref{matching}) take the form \bea
S_{\mu\nu}-\frac{1}{2}Sg_{\mu\nu}=c_{1}G_{\mu\nu}+c_{2}g_{\mu\nu}+c_{3}T_{\mu\nu},
\label{strange1} \eea ($S=S_{\mu\nu}g^{\mu\nu}$) with $c$'s being
the appropriate coefficients, while the contraction of the
generalized Gauss-Codazzi equation (\ref{GC}) gives \bea
S_{\mu\nu}=R_{\mu\nu}-\mathcal{R}^{\lambda}_{\,\,\,\mu\lambda\nu}.
\label{strange2} \eea From equations (\ref{strange1}),
(\ref{strange2}) we get \bea
\mathcal{R}^{\lambda\mu}_{\,\,\,\,\,\,\lambda\nu}-\frac{1}{2}\mathcal{R}^{\lambda\sigma}_{\,\,\,\,\,\,
\lambda\sigma}\delta^{\mu}_{\nu}=(1\!-\!c_{1})G^{\mu}_{\nu}-c_{2}\delta^{\mu}_{\nu}-c_{3}T^{\mu}_{\nu}.
\label{strange3} \eea Considering the $\mu$-covariant derivative
with respect to $|$ of equation (\ref{strange3}), transforming the
$|$ derivatives on the left hand side into $;$ derivatives, and
making use of the Bianchi identity, we arrive at equation
(\ref{con}). Note that in equation (\ref{strange3}) the extrinsic
curvatures have been eliminated, and the effective geometry is
directly connected to the bulk geometry. Finally, making use of
equations (\ref{matching}), (\ref{christmas}), (\ref{con}), we
obtain the exact conservation equation on the brane \bea
T^{\mu}_{\nu|\mu}=0, \label{final} \eea independently of the energy
momentum tensor in the bulk. This is contrary to the
five-dimensional case, where equation (\ref{final}) is valid only if
the mixed transverse-parallel components of the bulk energy-momentum
tensor vanish.

In conclusion, we have considered a codimension two (thin)
braneworld in Einstein-Gauss-Bonnet (-induced gravity) theory, where
the addition of the Gauss-Bonnet term is known to make meaningful
the situation when non-trivial braneworld matter content is
included. Using appropriate components of the field equations, and
the generalized Gauss-Codazzi-Mainardi equations holding for higher
codimensions, we show that the energy-momentum of the brane is
always conserved, independently of the bulk energy momentum tensor,
or a possibly variable deficit angle.

\[ \]
{\bf Acknowlegements} We wish to thank C. Charmousis and J. Garriga
for useful discussions. This work is supported by a European
Commission Marie Curie Fellowship, under contract
HMPF-CT-2004-400737.

\end{document}